\documentclass[prl,aps,preprintnumbers,twocolumn]{revtex4}
\usepackage{epsfig}
\usepackage{amsmath}
\usepackage{mathrsfs}

\makeatletter
\def\slash#1{{\mathpalette\c@ncel{#1}}} 
\makeatother

\newcommand\beq{\begin{eqnarray}}
\newcommand\eeq{\end{eqnarray}}

\newcommand\ra{\rangle}

\newcommand{\Dcal}{{\mathscr{D}}}

\def\ellslash{\rlap/{\mkern-1mu \ell}}
\def\Hcal{{\cal{H}}}

\def\Sslash{\rlap/{\mkern-1mu S}}
\def\pslash{\rlap/{\mkern-1mu p}}
\def\phatslash{\rlap/{\mkern-1mu \hat{p}}}
\def\Phatslash{\rlap/{\mkern-1mu {\hat{P}_h}}}
\def\Pslash{\rlap/{\mkern-1mu P}}

\def\ellslash{\rlap/{\mkern-1mu \ell}}
\def\kslash{\slash{\mkern-1mu k}}
\def\nslash{\slash{\mkern-1mu n}}

\def\Sslash{\slash{\mkern-1mu S}}

\begin{document}

\title{Universal Structure of Twist-3 Soft-Gluon-Pole
Cross Sections\\
for Single Transverse-Spin Asymmetry}

\author{Yuji Koike$^{1}$ and Kazuhiro Tanaka$^2$}
\affiliation{${}^1$Department of Physics, Niigata University,
Ikarashi, Niigata 950-2181, Japan\\
${}^2$Department of Physics, Juntendo University, Inba-gun, Chiba
270-1695, Japan}
\date{\today}

\begin{abstract}
We prove that twist-3 soft-gluon-pole (SGP) cross section for single
spin asymmetries (SSA) is determined by a certain ``primordial'' twist-2 cross section
up to kinematic and 
color factors in the leading order perturbative QCD. 
In particular, 
for the processes in which the partonic hard scattering occurs among
massless partons,
the invariance of the ``primordial'' partonic cross section under scale
transformation
leads to remarkable simplification of the SGP cross section, reproducing
compact form 
that was recently observed for pion production $p^\uparrow p\to \pi X$ 
and direct-photon production $p^\uparrow p\to \gamma X$.
\end{abstract}

\maketitle

There has been growing interest in the single transverse spin asymmetry
(SSA)
in high-energy semi-inclusive reactions (see \cite{Review} for a
review).
Due to the ``naively $T$-odd'' nature of
SSA, it occurs as an interference between the
amplitudes which have different phases, with one of 
the amplitudes 
causing 
single helicity-flip in the scattering.  
In the region of the large transverse momentum of the observed particle
in the final state,
the approach based on the collinear factorization in perturbative QCD
becomes
valid, and the SSA can be described as a twist-3 observable in this
region\,\cite{ET82,QS91,ekt06}.  
In our recent paper\,\cite{ekt06} we 
have established the formalsim of the twist-3 mechanism for SSA,
showing the factorization and gauge invariance 
of the single spin-dependent cross section in the lowest order
perturbative QCD, which 
has given a solid theoretical basis to the previously obtained 
cross section formulae for
SSA\,\cite{QS91,QS99,KK00,KK01,EKT,JQVY06,JQVY06DIS,KQVY06}. 
The connection of this mechanism to another approach for SSA
based on so-called ``$T$-odd'' distribution/fragmentation functions with
parton's 
intrinsic $k_\perp$\,\cite{Sivers} 
has also been studied recently\,\cite{JQVY06,JQVY06DIS}.

In the twist-3 
mechanism for SSA, the interfering phase is provided by
the
pole of an internal propagator of the partonic hard cross sections, and
those
poles are classified as soft-gluon-pole (SGP), soft-fermion-pole (SFP)
and
hard-pole (HP), depending on the parton's
momentum fraction
at the
poles in
the twist-3 quark-gluon correlation function.
In our recent paper\,\cite{KT06}, we have shown that
the hard cross section from SGP
is completely
determinend by the corresponding 
twist-2 unpolarized cross section,
for semi-inclusive deep inelastic scattering
(SIDIS),
$ep^\uparrow\to e\pi X$, and
Drell-Yan and direct-photon production, $p^\uparrow p\to\gamma^{(*)}X$.
In the present paper, we shall extend this study to SSA in ``QCD-induced'' $pp$
collisions,
such as $p^\uparrow p\to \pi X$ and $pp\to \Lambda^\uparrow X$, and will
show that the corresponding SGP contribution also obeys similar but more sophisticated pattern, 
i.e.,
can be expressed as a derivative
of a 
certain twist-2 cross section 
with respect to the parton's momentum
originating from initial- or final-state hadron,
up to color factors.
In addition, 
we shall show 
that the scale 
invariance of the relevant Born
cross sections among massless partons
leads to a
remarkably compact formula for the
SGP cross section as was the case for the direct-photon 
production\,\cite{KT06}.  
This clarifies the origin of the observed compact formula for the SGP
cross section
for $p^\uparrow p\to\pi X$\,\cite{KQVY06}.

To be specific we consider SSA for $p^\uparrow p\to \pi X$, in
particular,
the contribution from twist-3 distribution function of the
transversely polarized nucleon, 
defined as\,\cite{ekt06,EKT} 
\begin{eqnarray}
\hspace{-0.4cm}
M^{\beta,a}_{Fij}&&\!\!\!\!\!\!\!\!\!\!\!\! (x_1,x_2)
=\int {d\lambda\over 2\pi}
{d\mu\over 2\pi}
e^{i\lambda x_1}e^{i\mu(x_2-x_1)}\nonumber\\
&&\;\;\;\;\;\;\;\times\  \langle p\ S_\perp |\bar{\psi}_j(0)
{gF^{\beta\rho}_{a}(\mu n)n_\rho}
\psi_i(\lambda n)|p\ S_\perp \rangle\nonumber\\
&=&\!\!
{M_N\over 4} {2 \over N_c^2-1} \left( \pslash t^a \right)_{ij} 
\epsilon^{\beta pnS_\perp}
{G_F^q(x_1,x_2)}+ \cdots,
\label{GF}
\end{eqnarray}
where $|p\ S_\perp\ra$ is the nucleon state with momentum $p^\mu$ ($p^2 =M_N^2$) and
spin vector $S_\perp^\mu$ 
($S_\perp^2 = -1$), 
$F^{\beta \rho}_a$ is the gluon field strength tensor with octet color index $a$,  
and spinor indices $i, j$ associated with both 
Dirac and color structure of the quark field
$\psi$ 
are shown explicitly.  
We suppressed the path-ordered gauge-links that connect the fields on the lightcone
and make (\ref{GF}) gauge invariant.
In the twist-3 accuracy, $p$ (and momenta of all other hadrons as
well) 
can be regarded as lightlike $p^2=0$ and
$n$ is another lightlike vector satisfying $p\cdot n=1$.  
Here we take $p$ in the +z
direction as  
$p^-=p_\perp=0$ and $n^+=n_\perp=0$.   
The twist-3 quark-gluon correlation function
$G_F^q(x_1,x_2)$ is dimensionless and satisfies $G_F^q(x_1,x_2)=G_F^q(x_2,x_1)$.
The ellipses in (\ref{GF}) stand for Lorentz structures 
associated with another twist-3 correlation function, 
antisymmetric under $x_1 \leftrightarrow x_2$\,\cite{EKT},
and with twist higher than three; 
the function vanishing at $x_1 =x_2$ is irrelevant (see below).

We first recall the structure of the familiar twist-2 unpolarized cross section for 
$p(p)+p(p')\to\pi(P_h)+ X$: 
\beq
&& \!\!\!\!\!\!\!\!\!\!\!\!\!\!\!
E_h{d^3\sigma^{\rm tw2}\over d^3 P_h}={\alpha_s^2 \over S}
\sum_{a,b,c=q,\bar{q},g}\int{dz\over z^2}
{dx'\over x'}
{dx\over x}
\nonumber\\
&&\qquad\;\;\;\times 
f_a(x) f_b(x')D_c(z)H_U^{ab,c},
\label{twist2}
\eeq
where $E_h=P_h^0$ and $S=(p+p')^2$.
$f_q(x)$ and $D_q(z)$ ($f_g(x)$ and $D_g(z)$) are the quark (gluon) 
distribution and fragmentation functions
for the proton and the pion, respectively, 
and these are convoluted 
with the corresponding partonic hard-scattering functions
$H_U^{ab,c}$, over the relevant momentum fractions $x$, $x'$ and $z$. 
We consider
the contribution from quark-quark scattering (Fig.~1) in detail; 
$q(xp)+q(x'p')\to q(P_h/z)+q(xp+x'p'-P_h/z)$.
The relevant hard-scattering function
can be written as 
\begin{equation}
H_U^{qq,q} = \frac{1}{2}{\cal C}_q^2  {\rm Tr} 
\left[ \Hcal^{(0)}(xp, x'p', P_h /z ) \frac{1}{2}x \pslash \right],
\label{hard0}
\end{equation}
where $\Hcal^{(0)}_{ji}(xp,x'p',P_h/z)$ stands for the $qq\to qq$
hard scattering part whose spinor indices $i,j$ are associated with the initial 
quark with momentum $xp$ and are traced (${\rm Tr}[\cdots]$) over both Dirac and color indices
with the insertion $x\pslash/2$ to project onto the unpolarized distribution
$f_q(x)$.
\begin{figure}[t]
\begin{center}
\includegraphics[height=3.0cm,clip]{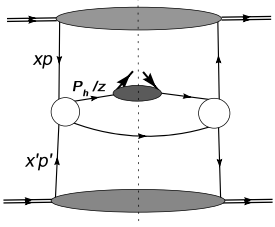}
\end{center}
\vskip -0.7cm \caption{\it Generic Feynman diagram contributing
to twist-2 unpolarized cross section for $p p\to \pi X$,
through the
``$q(xp)+q(x'p')\to q(P_h/z)+q(xp+x'p'-P_h/z)$'' scattering channel.
White circles denote the hard scattering between partons.}
\end{figure}
The factor ${\cal C}_q\equiv 1/N_c$ is the color-averaging factor for the initial quarks;
${\cal C}_q \rightarrow  {\cal C}_g\equiv 1/(N_c^2-1)$
when we consider the gluon initiated processes associated with the gluon distribution
$f_g(x')$.

The hard part (\ref{hard0})
can be explicitly written as 
(here and below, we frequently use the shorthand notation for the partonic momenta, 
$\hat{p} \equiv xp$, $\hat{p}' \equiv x'p'$ and $\hat{P}_h \equiv P_h/z$)
\beq
&&\!\!\!\!\!\!\!
\Hcal_{ji}^{(0)}(\hat{p},\hat{p}', \hat{P}_h )=\! \sum_{m,m',r,r',s,s'}
\!\!\! \bar{\cal F}_{jm',s'r'}( \hat{p},\hat{p}',\hat{P}_h )
\left( \Phatslash\right)_{m'm}\nonumber\\
&&\qquad\quad \times\Dcal_{r'r}(\hat{p}+\hat{p}'-\hat{P}_h)
{\cal F}_{mi,rs}( \hat{p},\hat{p}',\hat{P}_h )
\left( \phatslash' \right)_{ss'}, 
\label{bornXsection}
\eeq
using the Born amplitude ${\cal F}_{mi,rs}( \hat{p},\hat{p}' , \hat{P}_h )$ for 2-to-2
scattering, 
``$q_i( \hat{p} )+q_s(\hat{p}') \rightarrow q_m ( \hat{P}_h )+q_r(\hat{p}+\hat{p}'-\hat{P}_h)$'',
with the factors for its external lines amputated: 
\beq
&& \!\!\!\!\!\!\!\!\!\!\!
{\cal F}_{mi,rs}(\hat{p},\hat{p}' , \hat{P}_h )=\left(i\gamma^\mu t^b \right)_{mi}
\left(i\gamma_\mu t^b \right)_{rs}
{-i\over (\hat{p}-\hat{P}_h )^2+i\varepsilon} 
\nonumber\\
&&\qquad\quad+\left(i\gamma^\mu t^b \right)_{ri}
\left(i\gamma_\mu t^b \right)_{ms}{-i\over (\hat{P}_h - \hat{p}' )^2+i\varepsilon}, 
\label{F}
\eeq
where $i\gamma^\mu t^b$ comes from each quark-gluon vertex, 
and we employ the Feynman gauge
for the gluon propagator. 
In (\ref{bornXsection}), the amplitude in the RHS of the cut is obtained by 
$\bar{\cal F} ( \hat{p},\hat{p}',\hat{P}_h )
=(\gamma^0\otimes \gamma^0) {\cal F}^\dagger(\hat{p},\hat{p}',\hat{P}_h )(\gamma^0\otimes \gamma^0)$,
where 
each $\gamma$-matrix structure in 
${\cal F}( \hat{p},\hat{p}',\hat{P}_h )$
is sandwitched by $\gamma^0$, after taking the hermitian conjugate denoted by ``$\dagger$''.  
$\Dcal_{r' r}( \hat{p}+\hat{p}'-\hat{P}_h )$ represents
the cut quark-line for the unobserved final parton carrying the
large transverse-momentum as
$\Dcal_{r'r}(k)=\sum_{\rm spins}u_{r'}(k)\bar{u}_r(k)\delta\left(k^2\right)=
\left(\kslash\right)_{r'r}
\delta\left(k^2\right)$,
where $u(k)$ is the spinor for a quark with momentum $k$, and 
the Dirac structures ${\phatslash^\prime}$ and $ {\Phatslash}$
are assciated with $f_q (x')$ and $D_q (z)$ of (\ref{twist2}) with $b=q, c=q$.

With this convention for the twist-2 cross section, one can proceed to
derive the SGP cross section.  
The formalism for the twist-3 calculation has been establised in
\cite{ekt06} in the context of SSA for SIDIS, 
and it is straightforward to
extend it to $p^\uparrow p\to\pi X$.  
The relevant hard part is derived from the collinear expansion of
a set of cut Feynman diagrams of the type of Fig.~2 which are obtained by
attaching the additional 
gluon, generated by twist-3 effect of (\ref{GF}),
to the 2-to-2 partonic Born subprocess in Fig.~1.
We denote the sum of the partonic subprocess of those diagrams
as $\Hcal_{ji}^{(1)\sigma,a} (k_1,k_2, \hat{p}', \hat{P}_h )$ 
analogously to $\Hcal^{(0)}_{ji}(\hat{p}, \hat{p}', \hat{P}_h )$ of (\ref{hard0}), where  
$\sigma$ and $a$ are the Lorentz and color indices associated with the additional gluon.  
Applying the collinear expansion to these diagrams\,\cite{ekt06}, 
one obtains the general formula for the twist-3 
contribution to the spin-dependent 
cross section as
\beq
&& \!\!\!\!\!\!\!\!
E_h{d^3\sigma \over d^3 P_h}={\alpha_s^2\over 2S} 
\int{dz\over z^2} 
{dx' dx_1 dx_2 \over x' } 
\, {\cal C}_q f_q(x') D_q(z)\nonumber\\
&& \!\!\!\!\!\!\!\! \times 
{\rm Tr} \left[i\omega^\alpha_{\ \,\beta}\!
\left.{\partial \Hcal^{(1)\sigma,a} (k_1,k_2, \hat{p}', \hat{P}_h ) p_\sigma \over \partial
k_{2\perp}^\alpha}\right|_{\rm c.l.} \!\!\! M_F^{\beta,a}(x_1,x_2) \right] ,
\label{SGP}
\eeq
where 
$M_F^{\beta,a}(x_1,x_2)$ is defined in (\ref{GF}), 
$\omega^\alpha_{\ \, \beta}=g^\alpha_{\ \beta}-p^\alpha n_\beta$, and
``c.l.'' indicates to take the collinear limit $k_{1,2} \rightarrow x_{1,2} p$.
We note that all other complicated terms arising in the collinear expansion
vanish due to Ward identities\,\cite{ekt06}.  
Our task is to identify and evaluate the SGP contribution in 
${\partial \Hcal^{(1)\sigma,a}(k_1,k_2,\hat{p}', \hat{P}_h ) p^\sigma / \partial
k_{2\perp}^\alpha} |_{\rm c.l.}$.

It is known that 
only the diagrams in Figs.~2(a) and (b),
corresponding to the final-state interaction (FSI) 
and the initial-state
interaction (ISI), respectively, 
survive as the SGP contributions,
while the other diagrams cancel out combined with 
the corresponding ``mirror'' diagrams \cite{QS99,KK00,KQVY06}. 
The contribution of these diagrams 
to $\Hcal^{(1)\sigma, a}(k_1,k_2, \hat{p}', \hat{P}_h )p_\sigma$ 
can be easily obtained from (\ref{bornXsection}) by a slight modification
in each factor.  
Noting that the attachment of the extra gluon
to the final parton 
can be implemented by modifying the factor $\Phatslash$ in (\ref{bornXsection}),
the contribution from 
Fig.~2(a) 
is given by 
\beq
&&\Hcal_F^{La}(k_1,k_2, \hat{p}', \hat{P}_h )=
\bar{\cal F}(k_2, \hat{p}', \hat{P}_h )
\nonumber\\
&&\quad\times
\left( {\Phatslash} \, i\pslash t^a\, 
{i\over
\kslash_1-\kslash_2+ \hat{\Pslash}_h + i\varepsilon}\right) 
\nonumber\\
&&\quad\times 
\Dcal(k_2+ \hat{p}' - \hat{P}_h ) 
{\cal F}(k_1,\hat{p}' , k_1-k_2+ \hat{P}_h )
\left({\phatslash^\prime} \right),
\label{SGPF}
\eeq
where the role of each element is defined as in (\ref{bornXsection}), and 
we have suppressed the spinor indices as well as the summation over those.   
The contribution from the mirror diagram of Fig.~2(a) is obtained similarly and is expressed as 
$\bar{\Hcal}_F^{La}(k_2,k_1,\hat{p}', \hat{P}_h )
=\gamma^0 \Hcal_F^{La\dagger}(k_2,k_1,\hat{p}', \hat{P}_h )\gamma^0$, so that the
total contribution from FSI is $\Hcal_F^{La}(k_1,k_2, \hat{p}', \hat{P}_h )+ 
\bar{\Hcal}_F^{La}(k_2,k_1,\hat{p}', \hat{P}_h ) \equiv \Hcal_F^{a}(k_1,k_2, \hat{p}', \hat{P}_h )$.
Likewise, ISI diagram in Fig.~2(b) gives 
\beq
&&\Hcal_I^{La}(k_1,k_2,\hat{p}',\hat{P}_h)=
\bar{\cal F}(k_2,\hat{p}',\hat{P}_h)\left(\hat{\Pslash}_h\right)\nonumber\\
& &\quad\times
\Dcal(k_2+\hat{p}'-\hat{P}_h)
{\cal F}(k_1,k_2 - k_1+\hat{p}' , \hat{P}_h)\nonumber\\
&&\quad\times
\left({i\over \kslash_2-\kslash_1+{\phatslash}'+ i\varepsilon}\,i\pslash\,t^a\,{\phatslash}'
\right). 
\label{SGPI}
\eeq
Combining this with its mirror contribution,
we denote the total contribution to 
$\Hcal^{(1)\sigma, a}(k_1,k_2, \hat{p}', \hat{P}_h )p_\sigma$ 
from ISI
as $\Hcal_I^{a}(k_1,k_2, \hat{p}', \hat{P}_h )$. 
Note that the quark propagators appearing in (\ref{SGPF}) and (\ref{SGPI}) have
a pole at $k_2 = k_1$, corresponding to the FSI and ISI SGPs, respectively, 
as indicated by a cross in Figs.~2(a) and (b).

\begin{figure}[t]
\begin{center}
\includegraphics[height=3.1cm,clip]{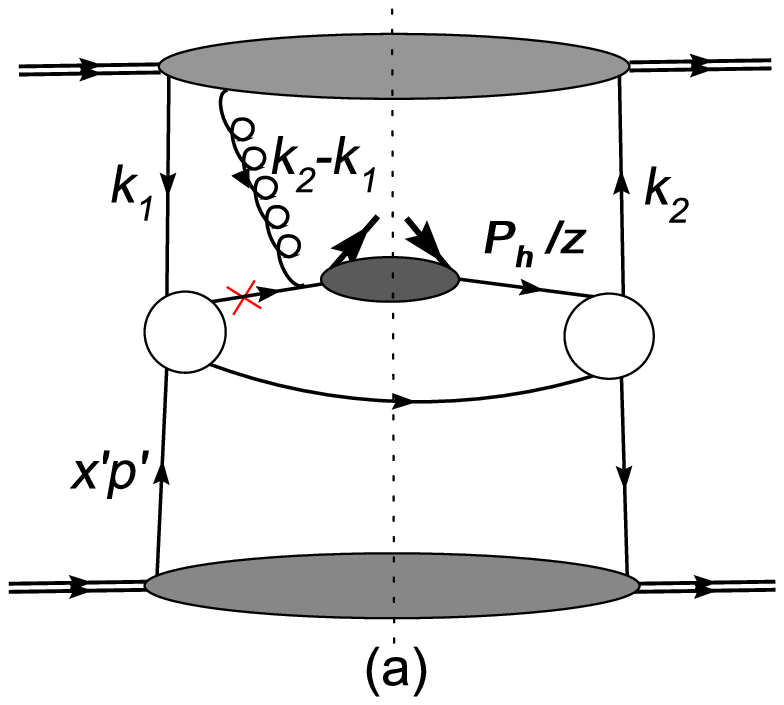}
\hspace{0.8cm}
\includegraphics[height=3.1cm,clip]{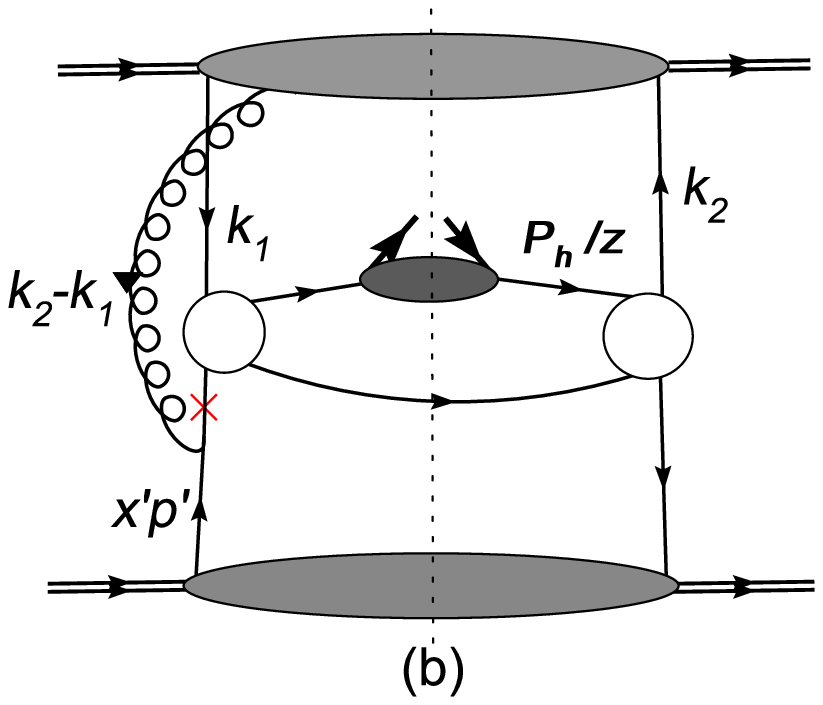}
\end{center}
\vskip -0.7cm \caption{\it Generic Feynman diagrams contributing
to SGP mechanism for SSA, through interference between the scatterings, $qqg \rightarrow qq$
and $qq \rightarrow qq$.
Mirror diagrams also contribute.}
\end{figure}

To obtain the SGP cross section, we substitute into (\ref{SGP}), 
$\Hcal^{(1)\sigma, a}(k_1,k_2, \hat{p}', \hat{P}_h )p_\sigma 
= \Hcal_F^{a}(k_1,k_2, \hat{p}', \hat{P}_h )+\Hcal_I^{a}(k_1,k_2, \hat{p}', \hat{P}_h )$. 
In calculating 
the derivative, $\partial/ \partial k_{2\perp}^\alpha$, 
we note the important feature of $\Hcal_{F,I}^a$ in (\ref{SGPF}) and (\ref{SGPI}): 
In $\Hcal_{F,I}^a$, 
the additional gluon couples
to an ``external parton'' associated with a final- or initial-state hadron, 
while the other part of $\Hcal_{F,I}^a$ is almost
the same as $\Hcal^{(0)}$ of (\ref{bornXsection}), except for the trivial ``momentum shift''
in the arguments of each component, ${\cal F}$, $\bar{\cal F}$, $\Dcal$.  
(This is in contrast to the cases for the HP and SFP contributions where
the extra gluon attaches to an internal parton line as well.)
Because of this structure, the attachment of the additional gluon
in $\partial\Hcal_{F,I}^a(k_1,k_2,\hat{p}',\hat{P}_h)/\partial k_{2\perp}^\alpha|_{\rm c.l.}$
can be systematically disentangled
by keeping the other factors in (\ref{SGPF}) and (\ref{SGPI})
intact, as was demonstrated in \cite{KT06} for SIDIS and Drell-Yan processes.  
It is straightforward to obtain the derivative as 
\beq
&& \!\!\!\!\!\!\!\!\!\!\!
\left.{\partial \Hcal^a_{Y}(k_1,k_2, \hat{p}', \hat{P}_h )\over \partial
k_{2\perp}^\alpha}\right|_{\rm c.l.}
\nonumber\\
&&\!\!\!\!\!\!
=\frac{1}{x_1 - x_2 + i \varepsilon}
{\partial \over \partial
\ell_{Y \perp}^\alpha}\left[ \
\left. \Hcal^{(0)}
(x_1p, \hat{p}', \hat{P}_h )\right|_{\ellslash_Y \rightarrow t^a \ellslash_Y} 
\right], 
\label{initial}
\eeq
with $Y=F, I$, and we set $\ell_{F}= \hat{P}_h$ and $\ell_{I}= - \hat{p}'$.
In the RHS of (\ref{initial}),
the form
$\ell_Y=-(\ell_{Y\perp}^2/2\ell_Y \cdot p)p+(\ell_Y \cdot p)n+\ell_{Y\perp}$ 
should be used for the on-shell momentum
$\ell_Y$, i.e., $\ell_Y^+$ should be treated as a dependent variable.  
The replacement ``$\ellslash_Y \rightarrow t^a \ellslash_Y$'', before taking this derivative,
indicates that we keep
$t^a$ in (\ref{SGPF}) and (\ref{SGPI}) intact, and 
such different insertion of color matrix results in 
different color factors
in (\ref{SGP}) for the FSI and ISI
contributions.

Inserting 
(\ref{initial})
into (\ref{SGP}) and evaluating the pole contribution, 
only $G_F^q(x_1, x_1)$ in 
(\ref{GF}) contributes.
The result gives the SGP cross section 
from the interference between $qqg\to qq$ and $qq\to qq$,   
but its hard cross section is now written as the
response of $qq \to qq$ scattering
to the change of the transverse momentum carried by the external parton, 
to which the extra gluon had coupled.
Also for other channels where antiquarks and/or gluons participate 
as initial- and final-state partons, 
it is straighforward to see that the results similar to (\ref{initial}) hold 
with appropriate 
substitutions of each factor in (\ref{bornXsection}),
${\cal F}$, $\Dcal$, ${\Phatslash}$ and ${\phatslash^\prime}$,
as was done for SIDIS and Drell-Yan processes in \cite{KT06}.   
Namely all these SGP contributions to (\ref{SGP}) can be expressed as the derivative of 
the corresponding 2-to-2 partonic Born subprocess, e.g.,
$q\bar{q}\to q\bar{q}$, $\bar{q}\bar{q}\to \bar{q}\bar{q}$,
$qg \to qg$, $q\bar{q}\to gg$, etc., which participates in the twist-2 unpolarized cross section 
(\ref{twist2}).
The only difference is the associated color structure; e.g., when the ``extra gluon'' couples to  
intial- or final-state gluon, $t^a$ in the RHS of (\ref{initial}) is simply replaced by its 
adjoint representation, $(t^a )_{bc} = -i f^{abc}$ \cite{KT06}.

A comment is in order regarding the gauge invariance of the SGP hard cross section
in (\ref{SGP}). 
Since (\ref{initial}) exhibits only the single pole at $x_1=x_2$,
the color gauge invariance with respect to the external gluon line is guaranteed by
$(x_2-x_1)\delta(x_1-x_2)=0$.
To make clear the gauge invariance regarding the internal gluon lines, 
we note that, in the twist-3 accuracy, 
one can replace the factor $\pslash$ from $M_F^{\beta,a}(x_1,x_2)$
in (\ref{SGP}) by $\kslash_1\nslash\kslash_2/(2x_1x_2)$
with the on-shell condition $k_i^2=0$ and perform 
the derivative $\partial/\partial k_{2\perp}^\alpha$ from the outside of
${\rm Tr}[\cdots]$, without changing the result\,\cite{ekt06}.
With this modification, all quark-gluon vertices in (\ref{SGPF}) and
(\ref{SGPI}) are sandwitched by the on-shell quark lines, and thus
the color gauge invariance is guaranteed for ${\rm Tr}[\cdots]$ in (\ref{SGP})
before taking the derivative by $\partial/\partial k_{2\perp}^\alpha$.  
The same argument applies to other channels as well.

Combinig all the above results,
one obtains the 
SGP cross section for SSA in $p^\uparrow p\to \pi X$, 
including all relevant channels.  
One should note
that the result
(\ref{initial}) holds in any frame with $p_{\perp}=0$;
in particular, 
one can move to a frame even with $\ell_{Y \perp}=0$ 
after taking the derivative.
In terms of the usual three partonic-invariants, 
$\hat{s}=(xp+ x'{p}')^2$,
$\hat{t}=(xp- {P}_h /z)^2$
and $\hat{u}=(x'p'- P_h /z)^2$, this implies that
the derivative $\partial/\partial \ell_{Y \perp}^\alpha$ 
can be performed through that for $\hat{u}$,
so that we obtain the SGP cross section as
\beq
&& \!\!\!\!\!\!\!\!
E_h{d^3\sigma^{\rm SGP}\over d^3 P_h}={\pi M_N\alpha_s^2 \over S}
\sum_{b,c=q,\bar{q}, g}\int{dz\over z^2}
{dx'\over x'}
{dx\over x}f_b(x') D_c(z)
\nonumber\\
&&\!\!\!\!\!
\times G_F^q (x,x)
\left\{ 
\left( {\hat{s}\over z \hat{t}} \epsilon^{S_\perp P_h pn} +
x' \epsilon^{S_\perp p' pn} \right)
{\partial H_F^{qb,c}(\hat{s},\hat{t},\hat{u}) \over \partial \hat{u}} \right.\nonumber\\ 
&&\;\;\; \left.- \left( \frac{1}{z}\epsilon^{S_\perp P_h pn} +
{x' \hat{t}\over\hat{s}} \epsilon^{S_\perp p' pn} \right)
{\partial H_I^{qb,c} (\hat{s},\hat{t},\hat{u})
\over \partial \hat{u}} \right\}, 
\label{SGPcross1}
\eeq
where 
\begin{equation}
H_{Y}^{qq,q}(\hat{s},\hat{t},\hat{u})= 
{\rm Tr} \left[ \frac{{\cal C}_q\, x \pslash\, t^a}{2 (N_c^2 -1)} \left. \Hcal^{(0)}
(\hat{p}, \hat{p}', \hat{P}_h )\right|_{\ellslash_Y \rightarrow t^a \ellslash_Y}  \right],
\label{hard3}
\end{equation}
with $\pslash t^a/[ 2(N_c^2 -1)]$ from (\ref{GF}), and similarly for other channels. 
The relation between (\ref{hard3}) and 
(\ref{hard0}) 
is apparent, and these are different only
in the ``color insertion'' of (\ref{hard3}).
Similar relation 
between $H_{F,I}^{qb,c}(\hat{s},\hat{t},\hat{u})$ in (\ref{SGPcross1})
and $H_U^{qb,c}(\hat{s},\hat{t},\hat{u})$ 
in (\ref{twist2}) also holds for each of all the other channels.   
To make the relation 
explicit,
we define $\hat{\sigma}_W^{qb,c}$ ($W=U, F, I$) as 
\begin{equation}
H_{W}^{qb,c}(\hat{s},\hat{t},\hat{u})= \hat{\sigma}_{W}^{qb,c} ( \hat{s},\hat{t},\hat{u})
\delta(\hat{s}+\hat{t}+\hat{u}). 
\label{phc}
\end{equation}
Explicit form of $\hat{\sigma}_{W}^{qb,c}$ is obtained as the sum of the contributions of  
2-to-2 Feynman diagrams for
the corresponding partonic subprocess, and they can be written as
$\hat{\sigma}_{W}^{qb,c}= \sum_{i}C^{qb,c}_{W(i)} \xi^{qb,c}_{(i)}$,
where $\xi^{qb,c}_{(i)}$ are dimensionless functions
expressed solely by $\hat{s},\hat{t}, \hat{u}$, and
$C^{qb,c}_{W(i)}$ denote the color factor associated 
with the diagram $i$.  
The difference among 
$\hat{\sigma}_{W}^{qb,c}$ is represented only through
the difference of $C^{qb,c}_{W(i)}$ among $W=U,F,I$.

Upon substitution of (\ref{phc})
into (\ref{SGPcross1}),
the derivative $\partial/\partial\hat{u}$ acts
both on
$\hat{\sigma}_{F,I}^{qb,c}$ and $\delta(\hat{s}+\hat{t}+\hat{u})$, the
latter contribution being handled by
partial integration with respect to $x$.   
One thus obtains the SGP cross section as
\beq
&& \!\!\!\!\!\!\!
E_h{d^3\sigma^{\rm SGP}\over d^3 P_h}={\pi M_N\alpha_s^2 \over S} 
\sum_{b,c=q,\bar{q},g}\int{dz\over z^2}
{dx'\over x'}
{dx\over x}f_b(x')D_c(z) 
\nonumber\\
&&\times \left[ x{d G_F^q(x,x) \over dx}-G_F^q(x,x)\right]
\left[ \frac{1}{z}\epsilon^{S_\perp P_h pn} +
{x' \hat{t}\over\hat{s}} \epsilon^{S_\perp p' pn} \right]
\nonumber\\
&& \;\;\;
\times \left( 
{\hat{s}\ \hat{\sigma}_F^{qb,c}(\hat{s},\hat{t},\hat{u})\over \hat{t}\hat{u}} 
-{\hat{\sigma}_I^{qb,c}(\hat{s},\hat{t},\hat{u}) \over \hat{u}}
\right)
\delta(\hat{s}+\hat{t}+\hat{u}).
\label{SGPformula}
\eeq
In a frame in which $p$ and $p'$ are collinear,
$\epsilon^{S_\perp p' pn }=0$.
We note that a straightforward calculation would produce additional
terms in the integrand, which are
proportional to $G_F^q (x,x) \left( \hat{s}{\partial / \partial\hat{s}}+
\hat{t}{\partial / \partial\hat{t}}+
\hat{u}{\partial /
\partial\hat{u}}\right)\hat{\sigma}_{F,I}^{qb,c}$.
However, such terms vanish, because of 
the scale-invariant property 
$\hat{\sigma}_{F,I}^{qb,c}(\hat{s},\hat{t},\hat{u})
=\hat{\sigma}_{F,I}^{qb,c}(\lambda\hat{s},\lambda\hat{t},\lambda\hat{u})$
as is obvious from the form of $\hat{\sigma}_{F,I}^{qb,c}$ 
discussed below (\ref{phc}).  
As a remarkable result, in (\ref{SGPformula}),
both ``derivative'' and ``non-derivative'' terms
of the SGP function $G_F^q(x,x)$
appear with the 
common partonic cross section expressed by $\hat{\sigma}_{F,I}^{qb,c}$, which are
identical to the twist-2
cross section up to color factors.  
Explicit forms of $\hat{\sigma}_{F,I}^{qb,c}$ obtained by ``color insertion''
to $\hat{\sigma}_{U}^{qb,c}$ coincide with the corresponding hard cross sections
derived in \cite{QS99,KK00,KQVY06}, 
and (\ref{SGPformula}) gives the complete
SGP cross section for $p^\uparrow p\to\pi X$. 
The above new derivation demonstrates
that the simplification as in (\ref{SGPformula}), which was recently observed in \cite{KQVY06},
arises from the quite general origin common to the SGP
contribution with twist-3 distributions:
(i) the general structure (\ref{initial}) obeyed by the hard part;
(ii) scale invariance
of the 2-to-2 partonic Born 
cross section among massless partons.
Because 
(i) and (ii) are independent of specific initial- or final-state,
the SGP contributions to SSAs in other processes with twist-3 distributions
follow the same pattern as (\ref{SGPcross1})
and (\ref{SGPformula}).  
For example, consider the
hyperon polarization in the unpolarized $pp$ collision, 
$pp\to\Lambda^\uparrow X$, in particular, 
the contribution from the chiral-odd twist-3
unpolarized distribution $E_F^q(x_1,x_2)$ combined with
the transversity fragmentation function $\delta\hat{q}(z)$
for $\Lambda^\uparrow$; $E_F^q(x,x)\otimes f_q(x')\otimes
\delta \widehat{q}(z)
\otimes \hat{\sigma}$\,\cite{KK01}.
The corresponding cross section is obtained by the replacement 
$\Pslash_h\to\gamma_5\Sslash_{h\perp}\Pslash_h$ in the above relevant formulae
to project onto $\delta\widehat{q}(z)$, where $S_{h\perp}$ is the transverse spin vector for
$\Lambda^\uparrow$, and 
by $\pslash \epsilon^{\beta pnS_\perp}G_F^q 
\rightarrow \gamma_5\pslash\gamma_\mu\epsilon^{\mu\beta np}E_F^q$
in (\ref{GF}) for the contribution of $E_F^q$.
The 
relevant 
twist-2 cross section is that for $p^\uparrow p\to\Lambda^\uparrow X$, i.e.,
involves (\ref{hard0}) with 
$\pslash\to\gamma_5\Sslash_\perp\pslash$.
It is clear that the similar relation between the SGP and the twist-2
cross sections holds for this case as well.  

The simplification of the SGP cross section due to the above (i) and (ii) 
can be also extended to the processes in which more number, $n$ $(\ge 3)$, of 
final-state (massless) partons
are involved.  This is because, for any such processes,
the hard part for the SGP cross section
is obtained by attaching the extra gluon from 
the twist-3 distribution function to the external parton lines
in the ``primordial'' twist-2 hard part, i.e., 
the 2-to-$n$ partonic Born subprocess.
Owing to this property, the relation similar to (\ref{initial}) 
holds, with $\ell_{Y}$ corresponding to the momenta of the relevant external partons. 
As the result, the hard part for the twist-3 SGP cross section 
associated with the $n$ final-state partons
is also completely determined
by the 
primordial twist-2 partonic 
cross section
up to color factors,
and   
the scale invariance of the twist-2 partonic cross section among massless partons
leads to further reduction of the formula of the SGP cross section.

The work of K.T. was 
supported by the Grant-in-Aid for Scientific Research No. C-16540266.

\end{document}